\documentclass[intlimits,twoside,a4paper]{article}
\usepackage{graphicx}
\usepackage{color}
\usepackage{colortbl}
\usepackage{wrapfig}
\usepackage[T2A]{fontenc}
\usepackage[utf8]{inputenc}
\usepackage[blocks]{authblk}
\usepackage{hyperref}
\usepackage{apacite}

\begin{document}

	\title{{Peculiarities of gender disambiguation and ordering of non-English authors' names for Economic papers beyond core databases}
		\thanks{This work was supported in part (OM) by the National Research Foundation of Ukraine, project No.~2020.01/0338.}
	}
	%\subtitle{Can we do this?}%\\ If so, write it here}
	
	%\titlerunning{}        % if too long for running head
	
%	
	\author{O. Mryglod}
	\affil{Institute for Condensed Matter Physics of the National Academy of Sciences of Ukraine, 1 Svientsitskii St., 79011 Lviv, Ukraine}
%		\email{olesya@icmp.lviv.ua}           

	\author{S. Nazarovets}
	\affil{Borys Grinchenko Kyiv University, 18/2 Bulvarno-Kudriavska Str., 04053 Kyiv, Ukraine}
%		\email{serhii.nazarovets@gmail.com}

	\author{S. Kozmenko}
	\affil{University of Social Sciences Spoleczna Akademia Nauk, 9 Sienkiewicza St., 90--113 {\L}\'{o}d\'{z}, Poland}
%		\email{kozmenko.uabs@gmail.com}

	\maketitle
	
	\begin{abstract}
		{This paper presents the results of further exploration of Crossref data related to Ukrainian Economics research} (the first part can be found in [Mryglod, O., Nazarovets, S. \& Kozmenko, S. (2021) Scientometrics, 126, 8187. \url{https://doi.org/10.1007/s11192-021-04079-7}]).\\
		\textbf{Purpose:} {To supplement the quantitative portrait of Ukrainian Economics discipline with the results of gender and author ordering analysis at the level of individual authors, special methods of working with bibliographic data with a predominant share of non-English authors are used. } The properties of gender mixing, the likelihood of male and female authors occupying the first position in the authorship list, as well as the arrangements of names are studied.\\
		\textbf{Design/methodology/approach:} 
		{A data set containing bibliographic records related to Ukrainian journal publications in the field of Economics is constructed using Crossref metadata. Partial semi-automatic disambiguation of authors' names is performed. First names, along with gender-specific ethnic surnames, are used for gender disambiguation required for further comparative gender analysis. Random reshuffling of data is used to determine the impact of gender correlations. 
			To assess the level of alphabetization for our data set, both Latin and Cyrillic versions of names are taken into account.}
		\textbf{Findings:} {
			The lack of well-structured metadata and the poor use of digital identifers lead to numerous problems with automatization of bibliographic data pre-processing, especially in the case of publications by non-Western authors. 
			The described stages for working with such specific data help to work at the level of authors and analyse, in particular, gender issues. Despite the larger number of female authors, gender equality is more likely to be reported at the individual level for the discipline of Ukrainian Economics.
			} 
		 The tendencies towards collaborative or solo-publications and gender mixing patterns are found to be dependent on the journal: the differences for publications indexed in Scopus and/or Web of Science databases are found. 

		 {It has also been found that Ukrainian Economics research is characterized by rather a non-alphabetical order of authors. }  \\
		\textbf{Research limitations:} Only partial authors' name disambiguation is performed in a semi-automatic way. Gender labels can be derived only for authors declared by full First names or gender-specific Last names.\\
		\textbf{Practical implications:} The typical features of Ukrainian Economic discipline can be used to perform a comparison with other countries and disciplines, to develop an informed-based assessment procedures at the national level. 
		{The proposed way of processing publication data can be borrowed to enrich metadata about other research disciplines, especially for non-English speaking countries. }
		\\
		\textbf{Originality/value:} To our knowledge, this is the first large-scale quantitative study of Ukrainian Economic discipline. 
		The results obtained are valuable not only at the national level, but also contribute to general knowledge about Economic research, gender issues and authors' names ordering. An example of the use of Crossref data is provided, while this data source is still less used due to a number of drawbacks. Here, for the first time, attention is drawn to the explicit use of the features of the Slavic authors' names.		 
	\end{abstract}
	
	%%%%%%%%%%%%%%%%%%%%%%%%%%%%%%%%%%%%%%%%%%%%%%%%%%%%%%%%%%%%%%%%%%%%%%%%%%%%%%%%%%%%%%%%%%%%%%%%%%%%%
	%\newpage
	\section{Introduction}
	%%%%%%%%%%%%%%%%%%%%%%%%%%%%%%%%%%%%%%%%%%%%%%%%%%%%%%%%%%%%%%%%%%%%%%%%%%%%%%%%%%%%%%%%
	{
		This paper contains the results that are part of a more general study  \cite{Mryglod2021}, the purpose of which is to perform a large-scale quantitative analysis of the Ukrainian Economics discipline using the publication data predominantly beyond Web of Science and Scopus databases, where currently more than 150 Ukrainian journals are indexed while the National List of recognized scientific journals\footnote{\url{https://mon.gov.ua/ua/nauka/nauka/atestaciya-kadriv-vishoyi-kvalifikaciyi/naukovi-fahovi-vidannya}} includes almost 1,500 titles.
		Motivated by the fact that Ukrainian research is still understudied because of its poor representation in core databases \cite{Aksnes2019} (and this is especially true for Social Sciences and Humanities -- SSH), we have made an attempt to provide a quantitative portrait of one of Ukrainian SSH disciplines using the Crossref database as an alternative data source.  
		Our interest in Ukraine is natural, as all three authors are Ukrainians and therefore motivated to contribute to a more transparent and evidence-based management of national research.
		Nevertheless, we also believe that this is an interesting case study that contributes to the better understanding of the research process in the developing countries of Eastern Europe, countries with a special historical heritage. Ukraine is characterised by non-English speaking and Cyrillic writing; this is especially true for the analysis of SSH. Economics is chosen as one of the most `visible' SSH disciplines, which is often considered as a transitional between the `hard' and `social' sciences \cite{Cainelli2012,Mryglod2012}.}
		
	{The quantitative analysis at the level of publications based on the Crossref data is rather straightforward. For example, an estimation of the number of authors per paper can be done even without sophisticated data pre-processing. However, the consideration of individual publication histories at the level of authors or gender analysis requires name disambiguation. 
		Obviously, this task is a big challenge, especially when dealing with non-Western names \cite{Treeratpituk2012,Gomide2017,Kim2021}. Since the majority of publications in the Ukrainian Economic discipline are related to local authors \cite{Mryglod2021}), it is natural to find mainly Ukrainian first and last names in our data set. 
		Although such peculiarities as the use of middle name or a prespecified order of parts in composite names \cite{Treeratpituk2012,Gomide2017} are not typical for Ukraine, a huge problem of transliteration (e.g., see also \cite{Muller2017})	 still exists. But there is another side of the coin: the so-called ethnicity can be used to improve gender disambiguation. Therefore, along with the initial motivation to contribute to the quantitative description of Ukrainian Economics discipline, a special emphasis is made on the methods of processing such specific bibliographic data. }
	
	{The research questions here are related both to the methods of data pre-processing and to the results of the analysis of these data:}
		\begin{itemize}
				
			\item[RQ1:] {What peculiarities of Ukrainian authors’ names have to be taken into account during the process of name and gender disambiguation?}
			\item[RQ2:] {What gender proportion is typical for the Ukrainian Economics discipline, and how can it be compared with similar results for other data sets (countries)?}
			\item[RQ3:] {What level of alphabetization characterizes the Ukrainian Economics discipline, and is it possible to identify any gender-related distinctions?	}
		\end{itemize}

	{Answering the main research questions, this paper serves also as another evidence of the usefulness of Crossref data as a potential source for bibliometric analysis. Economic publications are considered in many other studies, where the data from Web of Science or Scopus databases are exploited, see \cite{Schlapfer2010,DiVaio2009,Zhao2016,Wei2018,Truc2021}. And this is reasonable in order to assess top-impact output and reveal the research front in Economics. However, if the rest of the entire picture is needed, the potential of other sources such as Crossref can be efficiently used. In this context, Ukraine has an advantage -- a special interface called Open Ukrainian Citation Index (OUCI) was developed a few years ago. It provides a possibility to extract structured Crossref metadata related to all journals published in Ukraine \cite{Cheberkus2019}. Moreover, all these journals are labeled by subject category according to the Ukrainian national classification scheme. 
		 While this data source is not as comprehensive as a national current research information system could be, it provides a unique opportunity to supplement knowledge about research output of Ukraine.}
		
	{The paper is organized as follows: the description of our data set is provided in Section~\ref{sec_data}; the applied name disambiguation procedure is described in Section~\ref{sec_disam}; Section~\ref{sec_gender} describes the peculiarities of gender disambiguation procedure for our data and contains the results of gender analysis. 
		The author name ordering for the Ukrainian Economics discipline is studied in Section~\ref{sec_order}; the final discussion can be found in the last Section.}
	
	%%%%%%%%%%%%%%%%%%%%%%%%%%%%%%%%%%%%%%%%%%%%%%%%%%%%%%%%%%%%%%%%%%%%%%%%%%%%%%%%%%%%%%%%

%----------------------------------------------
    
    \section{Our data}\label{sec_data}
    This paper is a continuation of authors' previous work \cite{Mryglod2021}, which analyses  Crossref data for publications in Ukrainian journal papers in the field of Economics. The same principle of collecting data is used here, i.e., Crossref publication records related to Ukrainian Economics journals. {A not-for-profit membership organization Crossref\footnote{\url{https://www.crossref.org/}} collects metadata for publications with registered DOI (Digital Object Identifier) numbers. 
    	Each record contains basic bibliometric elements required for DOI registration (i.e., title, publication dates, authors, source title, volume and issue number, etc.). In addition, Crossref encourages its depositors to enrich metadata with authors’ affiliations, ORCID numbers, abstracts, lists of references, funding information, etc. These metadata are publically open, license-free and distributed through Crossref tools and APIs. Since 2018, DOI registration is required for any research paper to be officially recognized in Ukraine\footnote{See the Order of the Ministry of Education and Science of Ukraine N32 (2018, January 15). \url{https://zakon.rada.gov.ua/laws/show/z0148-18}}, therefore, Crossref metadata can be considered as a usefull source of information about the published outputs related to Ukrainian research. Moreover, a special web-interface -- Open Ukrainian Citation Index (OUCI)\footnote{\url{https://ouci.dntb.gov.ua/en/}} -- was developed to efficiently import these data. In particular, a number of search filters allowing classification of journals by their specialty are implemented into OUCI. In addition, information about current indexing of each journal in Scopus and Web of Science is provided.}

    The topical relevance of each journal is defined using the ``Speciality'' search filter. The following  specialities are considered to be related to the Economics field (a similar subject classification is used by the State Attestation Commission of Ukraine\footnote{{Official web-page of the State Attestation Commission of Ukraine: \url{https://mon.gov.ua/ua/tag/atestatsiya-kadriv-vishchoi-kvalifikatsii}}}: Economics; Tax and Accounting Policy; Finance, Banking and Insurance; Management; Marketing; Business, Entrepreneurship and Stock Markets; Public Administration; and International Economic Relations. Only journals in the National List of recognized scientific journals are considered. To exclude multidisciplinary editions, journal disciplines (upper classification level) are limited to the following list: Social and Behavioral Sciences; Management and Administration; Public Management and Administration; International Relations. 
    
     {In addition to the data available from Scopus and Web of Science, Crossref provides} an important piece of the puzzle required to build the full picture of Ukrainian Economics research. The results presented in this paper are based on the updated data set: data collection is performed at the end of February 2021. Altogether, 25,933 records for {papers published} in Ukrainian Economics journals between 2002 and 2020 were collected (the annual publication statistics is low before 2012 and rapidly increases afterwards: 97\% of records correspond to the period 2013--2020). {The imported records contain the following fields: Publication year; Journal ISSN number; DOI; Publisher; Title; Authors' names; Number of DOI-to-DOI citations (if information is provided by Crossref depositors); Journal is indexed in Scopus Yes/No; Journal is indexed in Web of Sciences Yes/No (up-to-date information in the last two fields is added by OUCI). }

	\section{{Name d}isambiguation problem}\label{sec_disam}
	
	While data analysis at the level of papers is performed in \cite{Mryglod2021}, many interesting questions can be put at the level of authors. To give an example, typical individual productivity or authors' collaboration patterns have to be known to set benchmarks for comparing, assessing or detecting examples of unusual publishing behaviour.  What is also important, authors' gender is typically (and in this work) inferred from the given names. Therefore, gender label cannot be assigned if only initials are specified instead of the full name. However, merging various records related to the same person, allows us to enlarge the statistics of papers with genderized authors. For example, gender can be defined for \textit{KAFKA S.} (?) merged with \textit{KAFKA SOFIYA} (Female). However, the widely-known problem of name disambiguation appears if unique digital identifiers are not commonly used. With only names, it is impossible to guarantee  that two identically written names correspond to the same person. The uncertainty is higher if only initials are used. But everything is even more complicated in the case of publications by authors who are not native English speakers. It is possible to find numerous alternative transliterations of Cyrillic names for the same authors in our data set. Moreover, speaking of Ukrainian names, one should take into account the tradition of ``translating'' given names, and sometimes even last names, into Russian. For example, an author \textit{Bosovskaya} can be also mentioned as \textit{Bosovska}; \textit{Orlovskaya -- Orlovska; Mostenskaya -- Mostenska}.  Many of the first names can be transliterated to English using Ukrainian or Russian Cyrillic versions; some of the most used are: \textit{Mykola -- Nikolay, Oleksandr -- Aleksandr, Kateryna -- Ekaterina, Olena -- Elena}. It can be instantly noted that the names in these pairs correspond to different initials: \textit{M -- N, O -- A, K -- E, O -- E}, respectively. The space of possible alternatives is also expanded by using different short versions of names: \textit{Olena $\rightarrow$ Lena, Oleksiy $\rightarrow$ Alex, Anastasiya $\rightarrow$ Nastya, Tetyana $\rightarrow$ Tanya}, etc. Sometimes, the same name can be written in many ways, each of them is automatically recognized as a separate name. Moreover, metadata for Ukrainian Economics journals can be deposited not only in English, but also in Ukrainian (Russian). Last but not least, inaccurate usage of Latin and Cyrillic alphabets is a problem. The homoglyphs -- letters that look the same on a screen but coded differently -- are used arbitrarily. After all, 40 versions of the name \textit{Eugen} are found in the data set. Taken together, all these peculiarities of metadata of Ukrainian (non-native English) publications complicate the process of disambiguating the names of authors. 
	
	Due to the numerous nuances listed above, it is too difficult to perform a full disambiguation procedure automatically. That can be done only partially and only in a semi-automatic way. The following criteria and approaches are used:
	\begin{itemize}
		\item Identical names found in different papers are considered as related to one person, since the limited data set that corresponds to a particular subject area is studied. {
			The assumption that there is a low probability of duplicated names within our data set is confirmed by manually checking randomly selected records.} Of course, exceptions are possible. Authors'  records with identical first and second names are separated if both appear in the same paper.
		
		\item The existence of common co-authors for two authors is considered as an argument to merge corresponding records.		

		\item A manually created list of Ukrainian given names together with ``synonymical'' forms (Latin and Cyrillic) was used to find candidates for merging\footnote{The list is available online, \url{https://doi.org/10.6084/m9.figshare.13580297}}. A gender label is initially assigned to each name (manually). A few examples to demonstrate the variety of names are shown below.
		\subitem{$\bullet$} 
		\textit{ALEKSEI; ALEKSEY; ALEKSII; ALEXEI; ALEXEJ; ALEXEY; AЛЕК\-СЕЙ; ALEKCEY; OLEKSEY; OLEKSII; OLEKSIY; OLEKSYI; OLEKSІI; OLEKSІY; OLEXIJ; OLEXIY; АLEXEI; АЛЕКСЕЙ; ОЛЕК\-СІЙ; ОLEKSII; ALEKSY; OLEKCII; ОLEXII; OLEXII; ALEXSEY; \linebreak OLEKSIJ; ОLEKSIY;} -- Male
		\subitem{$\bullet$} 
		\textit{CHRISTINA; CHRISTINE; CHRYSTYNA; CRISTINA; HRIS\-TI\-NE; KHRYS\-TY\-NA; KRISTINA; KRISTINE; KRISTYNA; KRYSTYNA; \linebreak ХРИС\-ТИН\-А; КРИС\-ТИ\-НА; КРІСТІНА; KHRISTINA; KRISZTINA}; -- Female	
		
		\item The records are merged if no contradictions appear. To give an example, all names from the following list: ``\textit{KAFKA S.М.; KAFKA SOFIYA; КАФКА С.М.; KAFKA S.M.; KAFKA SOFIIA; KAFKA S.}'' are merged to get a single author's record. But an ambiguity exists for authors from the  list ``\textit{VERHUN А.; VERGUN ANDRIJ IVANOVYCH; VERHUN A.; VERHUN ANTONINA; VERHUN ANDRIJ}'': \textit{VERHUN A.}\footnote{It seems like the same name appears twice, since Cyrillic `\textit{А}' and Latin `\textit{A}' are used for the author's initial.} corresponds either to \textit{ANDRIJ} or to \textit{ANTONINA} -- no merging is performed in this case. 
	\end{itemize}

The list of authors' names was processed using the own Python code to find the list of candidates for merging and to mark them as more or less probable. The final merging was manually confirmed using the results of this preliminary automatic procedure. Additional manual checks were performed for particular cases, where candidates are considered as important ``players'' due to a large number of publications or co-authors. Merging was not performed for the pairs where ambiguity remains, but even so, the initial set of 31.5 thousand authors' records was reduced to 23,094. 
	
\section{Gender {disambiguation and} analysis}\label{sec_gender}

As mentioned before, the gender label for an author is inferred from his/her given name. Since the majority of authors are from Ukraine \cite{Mryglod2021}, Slavic first names are predominantly found in our data set. Besides the list of Slavic names manually labeled by gender, free web resource Genderize\footnote{Genderize.io | Determine the gender of a first name. \url{https://genderize.io/\#overview}} was partially used to detect gender for non-Slavic names\footnote{Only results for the names occurring at least 10 times with a probability of more than 0.9 were taken into account.}. Thus,  54.5\% of 23,094 author records were marked by gender: 7,748 (33.5\%) females and 4,865 (21\%) males. According to this, females appear in our data set approximately 1.59 times more often. 

Some typical endings of Slavic surnames can be considered as gender-specific. Author records genderized on the previous step were used as a validation subset in order to check whether surnames' endings are distinctive enough for our data set. The gender of 19.85\% females was repeatedly recognized using gender-specific endings of last names: \textit{``OVA'', ``EVA'', ``ОВА'', ``ЕВА'', ``АЯ'', ``AYA'', ``AIA'', ``INA'', ``ІНА''}. In their turn, 18.19\% of males were recognized using the following list of endings: \textit{``OV'', ``EV'', ``ОВ'', ``ЕВ'', ``ЄВ'', ``YI'', ``YJ'', ``KY'', ``KII'', ``KIJ''}. As it can be seen, such an approach is almost equally accurate for both genders: males are slightly less recognizable. Only 8 (0.1\%) female and 7 (0.14\%) male authors are re-marked incorrectly. Such a few examples can be considered as exceptions: male surnames with female-like endings \textit{DIBROVA, СОВА (SOVA), ELLAIA, JAYA, GLOVA, GECHBAIA, BOVA} and female surnames with male-like endings \textit{KYI, MYSHELOV, GLAMBOSKY, LAZANYI, BOKII, URSAKII, SUSHYI, IVANOV}. Although gender can be assigned incorrectly to a particular person, we believe that the results are statistically correct. 

Thus, gender labels were inferred from surnames for an additional 1,260 female and 828 male authors. Altogether, we continue with 63.7\% genderized authors' names. And having in mind that the number of male authors is slightly underestimated, one can state that 1.5 times more female authors are found. 

Our finding is in line with the statement in \cite{Larivire2013}, where Ukraine is mentioned among other ``countries with lower scientific output'' that are characterized by more prevalent female authorship. But our research is not cross-disciplinary, it is initially related to the Economics area. Moreover, while Web of Science data were used in \cite{Larivire2013}, we exploit Crossref as a data source in our work.  The remaining question is how different can be results obtained for top journals indexed in authoritative databases such as Scopus or Web of Science and for data beyond these sources. 

It is shown that gender disparities are disciplinary-dependent \cite{DeNicola2021}. The Economics discipline is considered rather as a male-dominated one\footnote{Gender in the global research landscape. \url{https://
		www.elsevier.com/\_\_data/assets/pdf\_file/0008/265661/ElsevierGenderReport\_final\_for-web.pdf}} \cite{Liu2020,Bayer2016}. For example, 
20.3\% female versus 63.4\% male authors were found for an economics-related data set analyzed  in \cite{Maddi2021}\footnote{The gender is undetectable for the rest authors.}. A similar proportion was reported in \cite{Liu2020}: ``The proportion of men is 2.45 times higher than that of women''. On the contrary, more female authors are found in our data\footnote{{Of course, one has to remember that different datasets are used in these different case studies.}}. Thus, it is natural to expect more papers from female authors (at least one author is recognized as female) -- 76.3\% than from male authors -- 48.4\%.
But let's look more deeply into the
individual contributions by female and male authors, as it was suggested in \cite{Huang2020}. The conclusion that ``female and male authors are largely indistinguishable when it comes to the number of publications per year'' supported also by results presented in \cite{Liu2020} is relevant to our data: authors of both genders publish approximately the same number of papers per year on average. To be more precise, 1.28 papers per year on average are published by male authors, and 1.34 by female authors.

\begin{figure}[h]
\vspace{-10mm}
\centerline{\includegraphics[width=0.65\textwidth]{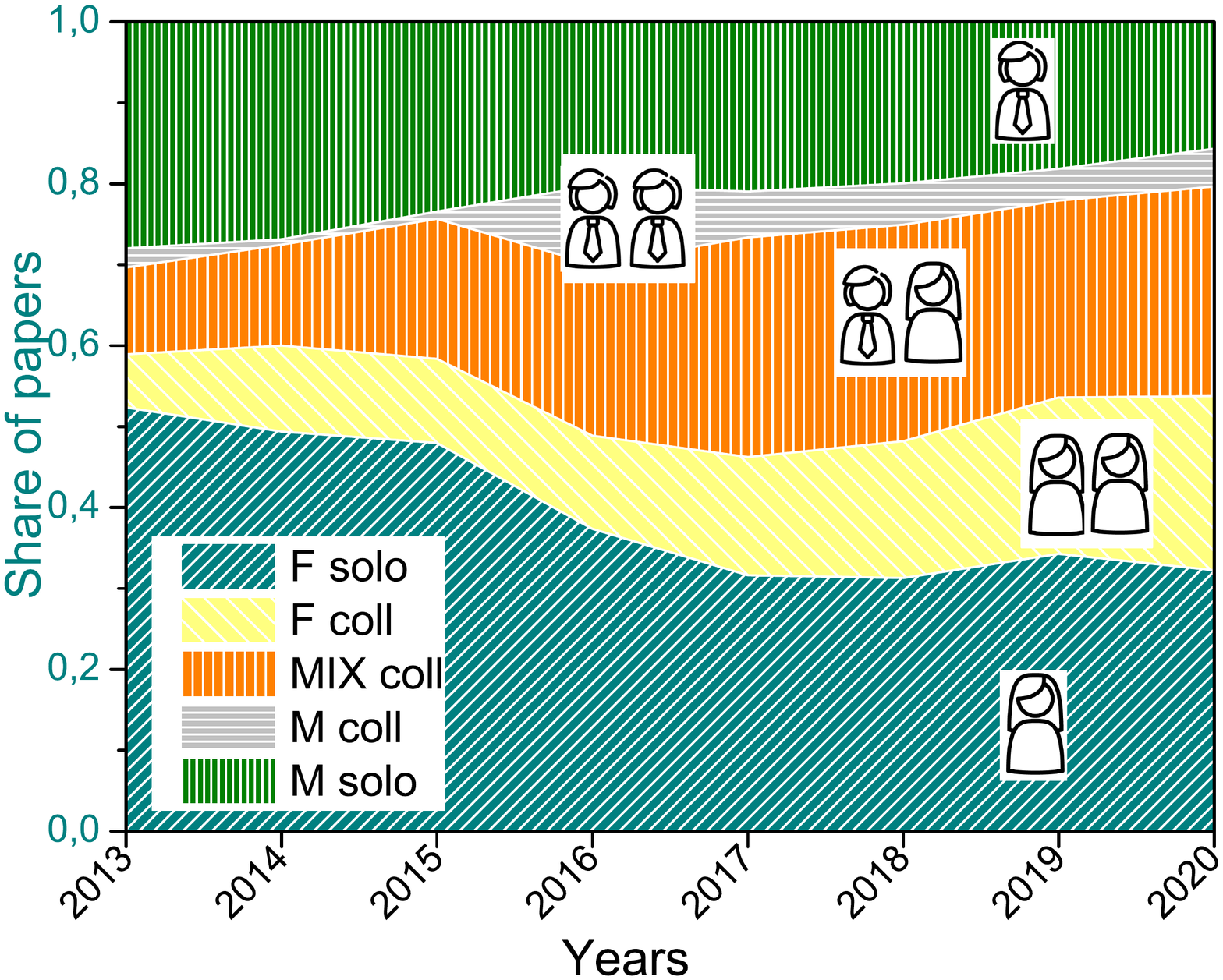}}
	\caption{Annual dynamics of shares of papers classified according to the gender of authors: a single female author (\emph{F solo}); two or more authors, only females (\emph{F coll}); two or more authors, at least one female and at least one male (\emph{MIX coll}); two or more authors, only males (\emph{M coll}); a single male author (\emph{M solo}). {The time window between 2013 and 2020 is chosen for visualization due to small annual statistics (less than 100 papers per year) before 2013.}}
	\label{Fig_GenderDyn}
\end{figure}

Another interesting relevant issue is the analysis of gender mixing and the patterns for forming authorship teams. The annual change of shares of papers classified according to gender of authors for our data is shown in Fig.~\ref{Fig_GenderDyn}. Five categories are defined here for 17,352 out of 25,933 papers (the rest cannot be classified due to the lack of gender information about authors): papers authored by a single person are labeled as \emph{F solo} or \emph{M solo} in correspondence to the author's gender (F stands for female and M for male); \emph{F coll} and \emph{M coll} labels are used only for papers where gender for all authors is defined and the same (solo-gender collaboration);  if not all authors are labeled by gender, but at least one female and one male are found, the paper is attributed to the \emph{MIX coll} category (cross-gender collaboration). Since not all authors in the data set are labeled by gender, the gender spectrum of papers can be considered as not conclusive. However, the same rule is applied here to data that correspond to different years, therefore, the tendencies are considered as informative. 

As Fig.~\ref{Fig_GenderDyn} shows, many papers correspond to the decreasing but still largest category of publications by a single female author. On the one hand, this is in line with the conclusion in \cite{Boschini2007}, where over-representation of single female authors is reported. On the other hand, it was already mentioned that the number of papers by females is expected to be larger in principle, simply due to the larger number of female authors. 
Indeed, for the same data set, if authors' gender labels `F', `M' and `undefined' are  randomly reshuffled first, the share of papers in the \emph{F solo} category $\approx 34.8\%$ is very close (even slightly larger) to the real value ($\approx 33.8\%$), see Table~\ref{Tab_gender_catgories}. 
Again, gender equality can be found at the individual level. The similar average share of solo-authored papers per author's portfolio is found for both genders (approximately 24\% if all genderized authors are taken into account; 28\% (F) and 30\% (M) if only genderized authors with at least 2 papers are considered). 
The similar conclusion is found in \cite{Kwiek2021} for Polish authors in the field of Economics -- the share of solo-publications in an individual portfolio is approximately equal for both male and female authors. According to this, a curious paradox can be observed: While the majority of papers in the Economics discipline are not collaborative \cite{Mryglod2021}, at the individual level, only each third (fourth if authors with a single paper are considered) paper is written without co-authors. No gender differences are also found in terms of the average number of authors per individual collaborative papers. 

The homophily of co-authorship groups with respect to authors' gender in the Economics field was discussed in \cite{Boschini2007}. At first glance, our results deny this conclusion. The cross-gender category is the second-largest one among the gender-labeled papers in our data, see Fig.~\ref{Fig_GenderDyn} and Table~\ref{Tab_gender_catgories}. However, it is easy to show that this share is smaller than expected. And indeed, solo-gender collaborative papers by female authors are over-represented in the real data set in comparison to reshuffled data. It is interesting to see that the shares of cross-gender papers and collaborative papers by male authors are remarkably higher for papers that are indexed in Scopus and/or WoS, see Table~\ref{Tab_gender_catgories}. To some extent, this is in agreement with the conclusion about the tendency to comparatively lower gender homophily in higher-impact journals in \cite{Holman2019}. Such gender mixing can be seen in a very positive way, since gender is one of the most important dimensions of team diversity, which, in its turn, is often considered as a powerful catalyst for creativity, see, e.g., \cite{Farhoomand2001,Reynolds2017,Liao2010} and references therein. 

\begin{table}[htb]
		\caption{Distribution of papers according to authors' gender: a single female author (\emph{F solo}); two or more authors, only females (\emph{F coll}); two or more authors, at least one female and at least one male (\emph{MIX coll}); two or more authors, only males (\emph{M coll}); a single male author (\emph{M solo}).}\label{Tab_gender_catgories}
		\begin{tabular}{|p{1.7cm}|p{0.9cm}|p{0.9cm}||p{1.5cm}||p{0.9cm}|p{0.9cm}|p{0.9cm}|p{0.9cm}|}
			\hline  & \multicolumn{2}{|c||}{\parbox{2cm}{Entire data set}}& \parbox{1.5cm}{Reshuffled$^*$}&
			\multicolumn{2}{|c|}{\parbox{2cm}{Not indexed in Scopus and/or WoS}}& \multicolumn{2}{|c|}{\parbox{2cm}{Indexed in Scopus and/or WoS}}\\
			\hline  & \# records  &  Share&  Share&\# records & Share& \# records & Share  \\
			\hline
			\hline
			All papers& 25933&-- &--&22683&--& 3250&--\\
			\hline
			All papers assigned to one of the categories& 17352 & 100\% &(13785 records on average) &15422&100\%& 1930&100\%\\
			\hline
			F solo& 5871&33.8\%& 34.8\% &5684&\cellcolor[gray]{0.8}36.9\%&187&\cellcolor[gray]{0.8}9.7\%\\
			\hline
			F coll&3085&17.8\%&10.4\%&2841&18.4\%&244&12.6\%\\
			\hline
            MIX coll&4283&24.7\%&28.5\%&3236&\cellcolor[gray]{0.8}21\%&1047&\cellcolor[gray]{0.8}54.3\%\\
			\hline
			M coll&816&4.7\%&4.1\%&598&3.9\%&218&11.3\%\\
			\hline
            M solo&3297&19\%&22.2\%&3063&19.8\%&234&12.1\%\\
			\hline    
		\end{tabular}
		\vspace{1mm}
		
		$^*$Averages for 10 versions of the original set of publications labeled by categories after random reshuffling of authors' gender labels are provided.
	\end{table}

\section{Authors ordering}\label{sec_order}
Another aspect of forming collaboration teams -- the ordering of authors  -- can be studied using our data. The position of the author's name in a list, which is ordered neither randomly nor alphabetically, can be considered as a basis for credit allocation. In this case, it is also reasonable to investigate correlation between authors' gender and their roles (i.e., positions in co-authorship lists). For example, the statistics of female first-authored journal articles are studied in \cite{Thelwall2020}.

Economics is often considered as one of the fields where alphabetization is common (see, e.g., \cite{Frandsen2010,Waltman2012,Levitt2013,Kuld2017}), although the alphabetization rate in the economy has declined somewhat over the past decade \cite{Wohlrabe2022}. Moreover, top Economic journals are characterized by the share of alphabetized articles that is even higher compared to other Economic journals (70\% vs. 60\%, correspondingly) \cite{Levitt2013,Kuld2017}. Such a way of ordering can be interpreted as a declaration of equal authors’ contribution. At the same time, the first position can still be perceived as special by external assessors. The so-called ``alphabetical discrimination'' is discussed in \cite{Einav2006,Kuld2017}: The staff members of a top U.S. economic department whose surnames start with letters from the first part of the alphabet are found to be more tenured. In some sense, this can be seen as a consequence of the Thomas theorem \cite{Bornmann2020}:  even a groundless consideration of the first author as the principal one can cause the further advantage in an academic career. Therefore, the following two questions are addressed further: (i) Is there any gender preference for the first position in a co-authorship list? and (ii) Can we state that the alphabetical ordering of authors is typical for the Ukrainian Economics discipline, in general?

\emph{Gender of first authors.} Due to the larger share of female authors, it is expected also to get a larger share of collaborative papers, where the first positions in co-authorship lists are occupied by female authors. {But it is interesting to calculate the probability of being on the first position in a collaborative list for each author. The individual publication records for both male and female authors with at least three collaborative papers are analysed for this purpose: the share of collaborative papers, where the given author is in the first position, are counted.  According to our results, the probability to occupy first position is equal to 0.48 for authors of both genders}. In this context, rather gender equality is found.

\emph{Alphabetization.} The alphabetical or non-alphabetical order of authors is determined by their last names. For the authors with identical last names, initials are considered. Since the majority of author names hypothetically are Slavs, their alphabetization can be performed in two ways: for their original names written in Cyrillic or for the corresponding versions in Latin. Therefore, the possible alphabetical order of each sequence of names is checked twice: for the names as they are in metadata and for their transliterations. To give an example, the names in the list \emph{``MARTYNENKO VALENTYNA; ZAMOTA IRINA''} are found to be \emph{ordered} by Latin alphabet (\emph{\textbf{M}ARTYNENKO; \textbf{Z}AMOTA}), but their transliterated versions are \emph{not ordered} by Cyrillic alphabet: (\emph{\textbf{М}АРТИНЕНКО; \textbf{З}АМОТА}).

	The more authors in the collaboration list, the less the probability of accidental alphabetical authorship is (see, e.g., \cite{Waltman2012,Kuld2017}). Since small co-authorship lists are dominant for the Economics discipline, in many cases one cannot be sure whether the names are sorted alphabetically intentionally or unintentionally. But non-alphabetical sorting is an unambiguous indicator of other priority scheme usage. Therefore, we count the share of papers with authors' names ordered neither by Latin nor Cyrillic alphabets. Corresponding numbers for different publication samples are provided in Table~\ref{Tab_order}. For example, it can be seen that in almost a half of all collaborative articles, authors are not sorted by the last names. Moreover, knowing the exact numbers of authors in the rest of the papers, one can suggest that another 21.3\% of the papers are sorted in alphabetic order accidentally\footnote{Considering all collaborative papers in the initial data set, one can find that the total number of alphabetically sorted papers is 6,641. The probability for a paper to be unintentionally ordered alphabetically depends on the number of authors $n$: $P(n)=1/n!$. Correspondingly, the number of papers that are intentionally alphabetized might be smaller: $5,194(1-1/2!)=2,597$ duo-authored papers; $1,284(1-1/3!)=1,070$  trio-authored papers; $135(1-1/4!)\approx 129$ quarto-authored papers; $\ldots$.} \cite{Waltman2012,Kuld2017}. 
	{Therefore, we conclude that the level of alphabetization of authors' names in Ukrainian Economic papers turned out} to be lower than it was reported for other publication sets. 
	Moreover, the share of non-alphabetized articles indexed in Scopus or Web of Science databases is even larger. One can only speculate about the reasons for such feature of Ukrainian Economics research. While the first position in the list of authors is not encouraged officially, still it is considered as more beneficial due to its greater visibility (it is a common practice to mention just the first author to refer to the co-authored publication) and its special perception within a number of disciplines. The name of the first author appears at the beginning of the reference.
\begin{table}[htb]
		\caption{The numbers of non-alphabetically ordered papers and estimated volume of potentially alphabetized papers (adjusted values) for different data samples.}\label{Tab_order}
		\begin{tabular}{|p{3cm}|p{2.5cm}|p{2.5cm}||p{2.5cm}|}
			\hline  &Number of papers (collaborative only)& Number (\%) of papers with definitely non-alphabetical ordering of authors' names& Adjusted number (\%) of papers that are potentially alphabetically ordered by intention\\
			\hline \hline 
			Entire data sets&13237&6596(49.8\%) &3824 (28.9\%)\\\hline\hline
            Papers marked by gender categories&8184&4057 (49.6\%)&2384 (29.1\%) \\\hline			
			Papers indexed in Scopus and/or WoS&	2592&1672 (64.5\%)&601 (23.2\%) \\\hline
			Papers NOT indexed in Scopus and/or WoS&	10645&4924 (46.3\%)& 3223 (30.3\%)\\\hline\hline
			Cross-gender papers&4283&2369 (55.3\%) &1169 (27.3\%)\\\hline
			Solo-gender papers&3901&1688 (43.3\%)&1215 (31.1\%)\\\hline			
		\end{tabular}
	\end{table}

\section{Discussion and conclusions}
The analysis of Ukrainian journals within the Economics discipline, started in the previous work \cite{Mryglod2021}, is continued in this study. {
	Revealing the typical features of this particular segment of scholar literature is important for solving many practical issues related to the development of assessment procedures at the national level.}
	 However, another goal of this work is to reinforce the call for complete and qualitative metadata. Crossref database is used here to describe one of SSH disciplines for poorly studied European countries.
%
%While Economics is considered as one of the %most visible Social disciplines, only small %share of journal papers by Ukrainian economists %is can be found in Scopus or Web of Science %databases. The large share of relevant outputs %are published in local journals and it is not %easy to analyse these publications. Crossref %database can be used in order to get metadata %related to journal papers not indexed in the %authoritative databases. Quantitative analysis %of such data allows one to describe the overall %publication dynamics within Ukrainian Economics %field, to draw the picture of authorship, to %get some information about authors' geography, %etc.
%
Publication metadata related to Ukrainian Economic journals are collected from the Crossref database.
An attempt was made to conduct an analysis with an emphasis on gender effects at the level of individual authors. However, the procedure for disambiguating authors' names can be done only partially. A number of peculiarities of processing author names related to the usage of Cyrillic and local traditions of parallel usage of different forms of names and even surnames are highlighted. Since the gender of an author is inferred from the full first name, even partial merging of authors' records allows one to increase the statistics of publications with authors labeled by gender. 
 Moreover, a manually created list of gender-specific endings for Slavic last names was used to enlarge the number of genderized authors. Altogether, 63.7\% of 23,094 author records were labeled by gender, and the number of female authors is found to be 1.5 times larger than male authors. This result contradicts with the statements about the masculinum nature of Economics research. Alternatively, female dominance in Ukrainian Economics research may be considered as a hint about its specific thematic spectrum. According to \cite{Thelwall2019}, keywords related to qualitative and exploratory methods are statistically associated with female scholar authors, while other keywords related to quantitative methods are more related to male authors. In some sense, such sensitivity of gender representation to the topic selection is in line with the conclusion in \cite{West2013}: considerable differences in this context were observed for different Economics subfields. This reinforces our previous conclusions about the specific patterns of collaborativeness in Ukrainian Economics research. Still, an important caution exists: 
 this study is one of the rare examples where data beyond internationally recognized databases is used. Therefore, the guess about different nature of locally-oriented and internationally-oriented topics chosen for  Economics research remains relevant.

Gender mixing is analysed to find the evidence that gender plays a role when forming collaboration teams. All papers labeled according to five gender-related categories (solo-publications by males; solo-publications by females; solo-gender collaborations of males; solo-gender collaborations of females; and cross-gender collaboration) are considered. While only one third\footnote{One fourth if authors with single papers are considered as well} of individual papers are written without coauthors and there is a tendency towards more collaborative papers \cite{Mryglod2021}, the share of solo-publications remains high. One third of all papers are found to be solo-publications authored by female authors. The corresponding share for male authors is slightly smaller than expected. {Finally, while the share of cross-gender teams is larger than the shares of solo-gender teams (see Fig.~\ref{Fig_GenderDyn}), the results compared with randomly reshuffled data indicate that the share of cross-gender teams is considerably higher than it can be expected only for publications in the journal indexed in Scopus and/or WoS, see Table~\ref{Tab_gender_catgories}. }

It is shown that the level of alphabetization of authors' names in Ukrainian Economic papers is comparatively low. This is especially true for articles indexed in Scopus or Web of Science databases. 

Interestingly, different results of gender mixing are found for papers published in journals  indexed in Scopus or Web of Science, compared to the rest of publications. Remarkably,  while the largest share of papers solo-authored by female authors is expected due to the greater general number of female authors, this category of papers indexed in the international databases is the least represented one. Most papers in internationally recognized journals are characterized by cross-gender collaboration. This can be seen as the manifestation of the so-called reactivity of the Ukrainian Economics discipline (see \cite{Aistleitner2019,Sasvri2019}). One can speculate about the adaptive publishing behaviour: a different publishing or even research strategy is chosen depending on  the level of recognition and audience of the target journal. The similar conclusion can be drawn for different shares of papers where authors are listed alphabetically. Interestingly, while a high level of alphabetization is found for the Economics discipline in general and even higher for Economic publications in  internationally recognized journals,  the opposite pattern is observed in Ukrainian Economics research.

To conclude, the results of another case study is presented. Besides the findings specifically related to Ukrainian research, some key aspects related to the processing of non-English metadata are highlighted. It is worth emphasizing once more that many complications become irrelevant if unique digital identifiers are commonly used.

\section*{Acknowledgements}
While the results reported in the paper are obtained before the {massive russian invasion to Ukraine}, \textcolor{yellow}{authors would like to thank} \textcolor{blue}{all Ukrainian defenders} for the possibility to finalise and publish this work. The authors thank anonymous peer reviewers for improving the paper.
%%%%%%%%%%%%%%%%%%%%%%%%%%%%%%%%%%%%%%%%%%%%%%%%


\begin{thebibliography}{}

\bibitem [\protect \citeauthoryear {%
Aistleitner%
, Kapeller%
\BCBL {}\ \BBA {} Steinerberger%
}{%
Aistleitner%
\ \protect \BOthers {.}}{%
{\protect \APACyear {2019}}%
}]{%
Aistleitner2019}
\APACinsertmetastar {%
Aistleitner2019}%
\begin{APACrefauthors}%
Aistleitner, M.%
, Kapeller, J.%
\BCBL {}\ \BBA {} Steinerberger, S.%
\end{APACrefauthors}%
\unskip\
\newblock
\APACrefYearMonthDay{2019}{{\APACmonth{12}}}{}.
\newblock
{\BBOQ}\APACrefatitle {Citation patterns in economics and beyond} {Citation
  patterns in economics and beyond}.{\BBCQ}
\newblock
\APACjournalVolNumPages{Science in Context}{32}{4}{361--380}.
\newblock
\begin{APACrefDOI} \doi{10.1017/s0269889720000022} \end{APACrefDOI}
\PrintBackRefs{\CurrentBib}

\bibitem [\protect \citeauthoryear {%
Aksnes%
\ \BBA {} Sivertsen%
}{%
Aksnes%
\ \BBA {} Sivertsen%
}{%
{\protect \APACyear {2019}}%
}]{%
Aksnes2019}
\APACinsertmetastar {%
Aksnes2019}%
\begin{APACrefauthors}%
Aksnes, D\BPBI W.%
\BCBT {}\ \BBA {} Sivertsen, G.%
\end{APACrefauthors}%
\unskip\
\newblock
\APACrefYearMonthDay{2019}{}{}.
\newblock
{\BBOQ}\APACrefatitle {A criteria-based assessment of the coverage of Scopus
  and Web of Science} {A criteria-based assessment of the coverage of scopus
  and web of science}.{\BBCQ}
\newblock
\APACjournalVolNumPages{Journal of Data and Information Science}{4}{1}{1--21}.
\PrintBackRefs{\CurrentBib}

\bibitem [\protect \citeauthoryear {%
Bayer%
\ \BBA {} Rouse%
}{%
Bayer%
\ \BBA {} Rouse%
}{%
{\protect \APACyear {2016}}%
}]{%
Bayer2016}
\APACinsertmetastar {%
Bayer2016}%
\begin{APACrefauthors}%
Bayer, A.%
\BCBT {}\ \BBA {} Rouse, C\BPBI E.%
\end{APACrefauthors}%
\unskip\
\newblock
\APACrefYearMonthDay{2016}{{\APACmonth{11}}}{}.
\newblock
{\BBOQ}\APACrefatitle {Diversity in the Economics Profession: A New Attack on
  an Old Problem} {Diversity in the economics profession: A new attack on an
  old problem}.{\BBCQ}
\newblock
\APACjournalVolNumPages{Journal of Economic Perspectives}{30}{4}{221--242}.
\newblock
\begin{APACrefDOI} \doi{10.1257/jep.30.4.221} \end{APACrefDOI}
\PrintBackRefs{\CurrentBib}

\bibitem [\protect \citeauthoryear {%
Bornmann%
\ \BBA {} Marx%
}{%
Bornmann%
\ \BBA {} Marx%
}{%
{\protect \APACyear {2020}}%
}]{%
Bornmann2020}
\APACinsertmetastar {%
Bornmann2020}%
\begin{APACrefauthors}%
Bornmann, L.%
\BCBT {}\ \BBA {} Marx, W.%
\end{APACrefauthors}%
\unskip\
\newblock
\APACrefYearMonthDay{2020}{{\APACmonth{02}}}{}.
\newblock
{\BBOQ}\APACrefatitle {Thomas theorem in research evaluation} {Thomas theorem
  in research evaluation}.{\BBCQ}
\newblock
\APACjournalVolNumPages{Scientometrics}{123}{1}{553--555}.
\newblock
\begin{APACrefDOI} \doi{10.1007/s11192-020-03389-6} \end{APACrefDOI}
\PrintBackRefs{\CurrentBib}

\bibitem [\protect \citeauthoryear {%
Boschini%
\ \BBA {} Sj\"{o}gren%
}{%
Boschini%
\ \BBA {} Sj\"{o}gren%
}{%
{\protect \APACyear {2007}}%
}]{%
Boschini2007}
\APACinsertmetastar {%
Boschini2007}%
\begin{APACrefauthors}%
Boschini, A.%
\BCBT {}\ \BBA {} Sj\"{o}gren, A.%
\end{APACrefauthors}%
\unskip\
\newblock
\APACrefYearMonthDay{2007}{{\APACmonth{04}}}{}.
\newblock
{\BBOQ}\APACrefatitle {Is Team Formation Gender Neutral? Evidence from
  Coauthorship Patterns} {Is team formation gender neutral? evidence from
  coauthorship patterns}.{\BBCQ}
\newblock
\APACjournalVolNumPages{Journal of Labor Economics}{25}{2}{325--365}.
\newblock
\begin{APACrefDOI} \doi{10.1086/510764} \end{APACrefDOI}
\PrintBackRefs{\CurrentBib}

\bibitem [\protect \citeauthoryear {%
Cainelli%
, Maggioni%
, Uberti%
\BCBL {}\ \BBA {} De~Felice%
}{%
Cainelli%
\ \protect \BOthers {.}}{%
{\protect \APACyear {2012}}%
}]{%
Cainelli2012}
\APACinsertmetastar {%
Cainelli2012}%
\begin{APACrefauthors}%
Cainelli, G.%
, Maggioni, M\BPBI A.%
, Uberti, T\BPBI E.%
\BCBL {}\ \BBA {} De~Felice, A.%
\end{APACrefauthors}%
\unskip\
\newblock
\APACrefYearMonthDay{2012}{}{}.
\newblock
{\BBOQ}\APACrefatitle {Co-authorship and productivity among Italian economists}
  {Co-authorship and productivity among italian economists}.{\BBCQ}
\newblock
\APACjournalVolNumPages{Applied Economics Letters}{19}{16}{1609--1613}.
\PrintBackRefs{\CurrentBib}

\bibitem [\protect \citeauthoryear {%
Cheberkus%
\ \BBA {} Nazarovets%
}{%
Cheberkus%
\ \BBA {} Nazarovets%
}{%
{\protect \APACyear {2019}}%
}]{%
Cheberkus2019}
\APACinsertmetastar {%
Cheberkus2019}%
\begin{APACrefauthors}%
Cheberkus, D.%
\BCBT {}\ \BBA {} Nazarovets, S.%
\end{APACrefauthors}%
\unskip\
\newblock
\APACrefYearMonthDay{2019}{{\APACmonth{11}}}{}.
\newblock
{\BBOQ}\APACrefatitle {Ukrainian open index maps local citations} {Ukrainian
  open index maps local citations}.{\BBCQ}
\newblock
\APACjournalVolNumPages{Nature}{575}{7784}{596--596}.
\newblock
\begin{APACrefDOI} \doi{10.1038/d41586-019-03662-6} \end{APACrefDOI}
\PrintBackRefs{\CurrentBib}

\bibitem [\protect \citeauthoryear {%
Einav%
\ \BBA {} Yariv%
}{%
Einav%
\ \BBA {} Yariv%
}{%
{\protect \APACyear {2006}}%
}]{%
Einav2006}
\APACinsertmetastar {%
Einav2006}%
\begin{APACrefauthors}%
Einav, L.%
\BCBT {}\ \BBA {} Yariv, L.%
\end{APACrefauthors}%
\unskip\
\newblock
\APACrefYearMonthDay{2006}{{\APACmonth{02}}}{}.
\newblock
{\BBOQ}\APACrefatitle {What{\textquotesingle}s in a Surname? The Effects of
  Surname Initials on Academic Success} {What{\textquotesingle}s in a surname?
  the effects of surname initials on academic success}.{\BBCQ}
\newblock
\APACjournalVolNumPages{Journal of Economic Perspectives}{20}{1}{175--188}.
\newblock
\begin{APACrefDOI} \doi{10.1257/089533006776526085} \end{APACrefDOI}
\PrintBackRefs{\CurrentBib}

\bibitem [\protect \citeauthoryear {%
Farhoomand%
\ \BBA {} Drury%
}{%
Farhoomand%
\ \BBA {} Drury%
}{%
{\protect \APACyear {2001}}%
}]{%
Farhoomand2001}
\APACinsertmetastar {%
Farhoomand2001}%
\begin{APACrefauthors}%
Farhoomand, A.%
\BCBT {}\ \BBA {} Drury, D\BPBI H.%
\end{APACrefauthors}%
\unskip\
\newblock
\APACrefYearMonthDay{2001}{}{}.
\newblock
{\BBOQ}\APACrefatitle {Diversity and Scientific Progress in the Information
  Systems Discipline} {Diversity and scientific progress in the information
  systems discipline}.{\BBCQ}
\newblock
\APACjournalVolNumPages{Communications of the Association for Information
  Systems}{5}{}{}.
\newblock
\begin{APACrefDOI} \doi{10.17705/1cais.00512} \end{APACrefDOI}
\PrintBackRefs{\CurrentBib}

\bibitem [\protect \citeauthoryear {%
Frandsen%
\ \BBA {} Nicolaisen%
}{%
Frandsen%
\ \BBA {} Nicolaisen%
}{%
{\protect \APACyear {2010}}%
}]{%
Frandsen2010}
\APACinsertmetastar {%
Frandsen2010}%
\begin{APACrefauthors}%
Frandsen, T\BPBI F.%
\BCBT {}\ \BBA {} Nicolaisen, J.%
\end{APACrefauthors}%
\unskip\
\newblock
\APACrefYearMonthDay{2010}{{\APACmonth{10}}}{}.
\newblock
{\BBOQ}\APACrefatitle {What is in a name? Credit assignment practices in
  different disciplines} {What is in a name? credit assignment practices in
  different disciplines}.{\BBCQ}
\newblock
\APACjournalVolNumPages{Journal of Informetrics}{4}{4}{608--617}.
\newblock
\begin{APACrefDOI} \doi{10.1016/j.joi.2010.06.010} \end{APACrefDOI}
\PrintBackRefs{\CurrentBib}

\bibitem [\protect \citeauthoryear {%
Gomide%
, Kling%
\BCBL {}\ \BBA {} Figueiredo%
}{%
Gomide%
\ \protect \BOthers {.}}{%
{\protect \APACyear {2017}}%
}]{%
Gomide2017}
\APACinsertmetastar {%
Gomide2017}%
\begin{APACrefauthors}%
Gomide, J.%
, Kling, H.%
\BCBL {}\ \BBA {} Figueiredo, D.%
\end{APACrefauthors}%
\unskip\
\newblock
\APACrefYearMonthDay{2017}{}{}.
\newblock
{\BBOQ}\APACrefatitle {Name usage pattern in the synonym ambiguity problem in
  bibliographic data} {Name usage pattern in the synonym ambiguity problem in
  bibliographic data}.{\BBCQ}
\newblock
\APACjournalVolNumPages{Scientometrics}{112}{2}{747--766}.
\PrintBackRefs{\CurrentBib}

\bibitem [\protect \citeauthoryear {%
Holman%
\ \BBA {} Morandin%
}{%
Holman%
\ \BBA {} Morandin%
}{%
{\protect \APACyear {2019}}%
}]{%
Holman2019}
\APACinsertmetastar {%
Holman2019}%
\begin{APACrefauthors}%
Holman, L.%
\BCBT {}\ \BBA {} Morandin, C.%
\end{APACrefauthors}%
\unskip\
\newblock
\APACrefYearMonthDay{2019}{{\APACmonth{04}}}{}.
\newblock
{\BBOQ}\APACrefatitle {Researchers collaborate with same-gendered colleagues
  more often than expected across the life sciences} {Researchers collaborate
  with same-gendered colleagues more often than expected across the life
  sciences}.{\BBCQ}
\newblock
\APACjournalVolNumPages{{PLOS} {ONE}}{14}{4}{e0216128}.
\newblock
\begin{APACrefDOI} \doi{10.1371/journal.pone.0216128} \end{APACrefDOI}
\PrintBackRefs{\CurrentBib}

\bibitem [\protect \citeauthoryear {%
Huang%
, Gates%
, Sinatra%
\BCBL {}\ \BBA {} Barab{\'{a}}si%
}{%
Huang%
\ \protect \BOthers {.}}{%
{\protect \APACyear {2020}}%
}]{%
Huang2020}
\APACinsertmetastar {%
Huang2020}%
\begin{APACrefauthors}%
Huang, J.%
, Gates, A\BPBI J.%
, Sinatra, R.%
\BCBL {}\ \BBA {} Barab{\'{a}}si, A\BHBI L.%
\end{APACrefauthors}%
\unskip\
\newblock
\APACrefYearMonthDay{2020}{{\APACmonth{02}}}{}.
\newblock
{\BBOQ}\APACrefatitle {Historical comparison of gender inequality in scientific
  careers across countries and disciplines} {Historical comparison of gender
  inequality in scientific careers across countries and disciplines}.{\BBCQ}
\newblock
\APACjournalVolNumPages{Proceedings of the National Academy of
  Sciences}{117}{9}{4609--4616}.
\newblock
\begin{APACrefDOI} \doi{10.1073/pnas.1914221117} \end{APACrefDOI}
\PrintBackRefs{\CurrentBib}

\bibitem [\protect \citeauthoryear {%
Kim%
, Kim%
\BCBL {}\ \BBA {} Owen-Smith%
}{%
Kim%
\ \protect \BOthers {.}}{%
{\protect \APACyear {2021}}%
}]{%
Kim2021}
\APACinsertmetastar {%
Kim2021}%
\begin{APACrefauthors}%
Kim, J.%
, Kim, J.%
\BCBL {}\ \BBA {} Owen-Smith, J.%
\end{APACrefauthors}%
\unskip\
\newblock
\APACrefYearMonthDay{2021}{}{}.
\newblock
{\BBOQ}\APACrefatitle {Ethnicity-based name partitioning for author name
  disambiguation using supervised machine learning} {Ethnicity-based name
  partitioning for author name disambiguation using supervised machine
  learning}.{\BBCQ}
\newblock
\APACjournalVolNumPages{Journal of the Association for Information Science and
  Technology}{72}{8}{979--994}.
\PrintBackRefs{\CurrentBib}

\bibitem [\protect \citeauthoryear {%
Kuld%
\ \BBA {} O'Hagan%
}{%
Kuld%
\ \BBA {} O'Hagan%
}{%
{\protect \APACyear {2017}}%
}]{%
Kuld2017}
\APACinsertmetastar {%
Kuld2017}%
\begin{APACrefauthors}%
Kuld, L.%
\BCBT {}\ \BBA {} O'Hagan, J.%
\end{APACrefauthors}%
\unskip\
\newblock
\APACrefYearMonthDay{2017}{{\APACmonth{11}}}{}.
\newblock
{\BBOQ}\APACrefatitle {Rise of multi-authored papers in economics: Demise of
  the `lone star' and why?} {Rise of multi-authored papers in economics: Demise
  of the `lone star' and why?}{\BBCQ}
\newblock
\APACjournalVolNumPages{Scientometrics}{114}{3}{1207--1225}.
\newblock
\begin{APACrefDOI} \doi{10.1007/s11192-017-2588-3} \end{APACrefDOI}
\PrintBackRefs{\CurrentBib}

\bibitem [\protect \citeauthoryear {%
Kwiek%
\ \BBA {} Roszka%
}{%
Kwiek%
\ \BBA {} Roszka%
}{%
{\protect \APACyear {2021}}%
}]{%
Kwiek2021}
\APACinsertmetastar {%
Kwiek2021}%
\begin{APACrefauthors}%
Kwiek, M.%
\BCBT {}\ \BBA {} Roszka, W.%
\end{APACrefauthors}%
\unskip\
\newblock
\APACrefYearMonthDay{2021}{{\APACmonth{08}}}{}.
\newblock
{\BBOQ}\APACrefatitle {Gender-based homophily in research: A large-scale study
  of man-woman collaboration} {Gender-based homophily in research: A
  large-scale study of man-woman collaboration}.{\BBCQ}
\newblock
\APACjournalVolNumPages{Journal of Informetrics}{15}{3}{101171}.
\newblock
\begin{APACrefDOI} \doi{10.1016/j.joi.2021.101171} \end{APACrefDOI}
\PrintBackRefs{\CurrentBib}

\bibitem [\protect \citeauthoryear {%
Larivi{\`{e}}re%
, Ni%
, Gingras%
, Cronin%
\BCBL {}\ \BBA {} Sugimoto%
}{%
Larivi{\`{e}}re%
\ \protect \BOthers {.}}{%
{\protect \APACyear {2013}}%
}]{%
Larivire2013}
\APACinsertmetastar {%
Larivire2013}%
\begin{APACrefauthors}%
Larivi{\`{e}}re, V.%
, Ni, C.%
, Gingras, Y.%
, Cronin, B.%
\BCBL {}\ \BBA {} Sugimoto, C\BPBI R.%
\end{APACrefauthors}%
\unskip\
\newblock
\APACrefYearMonthDay{2013}{{\APACmonth{12}}}{}.
\newblock
{\BBOQ}\APACrefatitle {Bibliometrics: Global gender disparities in science}
  {Bibliometrics: Global gender disparities in science}.{\BBCQ}
\newblock
\APACjournalVolNumPages{Nature}{504}{7479}{211--213}.
\newblock
\begin{APACrefDOI} \doi{10.1038/504211a} \end{APACrefDOI}
\PrintBackRefs{\CurrentBib}

\bibitem [\protect \citeauthoryear {%
Levitt%
\ \BBA {} Thelwall%
}{%
Levitt%
\ \BBA {} Thelwall%
}{%
{\protect \APACyear {2013}}%
}]{%
Levitt2013}
\APACinsertmetastar {%
Levitt2013}%
\begin{APACrefauthors}%
Levitt, J\BPBI M.%
\BCBT {}\ \BBA {} Thelwall, M.%
\end{APACrefauthors}%
\unskip\
\newblock
\APACrefYearMonthDay{2013}{{\APACmonth{07}}}{}.
\newblock
{\BBOQ}\APACrefatitle {Alphabetization and the skewing of first authorship
  towards last names early in the alphabet} {Alphabetization and the skewing of
  first authorship towards last names early in the alphabet}.{\BBCQ}
\newblock
\APACjournalVolNumPages{Journal of Informetrics}{7}{3}{575--582}.
\newblock
\begin{APACrefDOI} \doi{10.1016/j.joi.2013.03.002} \end{APACrefDOI}
\PrintBackRefs{\CurrentBib}

\bibitem [\protect \citeauthoryear {%
Liao%
}{%
Liao%
}{%
{\protect \APACyear {2010}}%
}]{%
Liao2010}
\APACinsertmetastar {%
Liao2010}%
\begin{APACrefauthors}%
Liao, C\BPBI H.%
\end{APACrefauthors}%
\unskip\
\newblock
\APACrefYearMonthDay{2010}{{\APACmonth{11}}}{}.
\newblock
{\BBOQ}\APACrefatitle {How to improve research quality? Examining the impacts
  of collaboration intensity and member diversity in collaboration networks}
  {How to improve research quality? examining the impacts of collaboration
  intensity and member diversity in collaboration networks}.{\BBCQ}
\newblock
\APACjournalVolNumPages{Scientometrics}{86}{3}{747--761}.
\newblock
\begin{APACrefDOI} \doi{10.1007/s11192-010-0309-2} \end{APACrefDOI}
\PrintBackRefs{\CurrentBib}

\bibitem [\protect \citeauthoryear {%
Liu%
, Song%
\BCBL {}\ \BBA {} Yang%
}{%
Liu%
\ \protect \BOthers {.}}{%
{\protect \APACyear {2020}}%
}]{%
Liu2020}
\APACinsertmetastar {%
Liu2020}%
\begin{APACrefauthors}%
Liu, J.%
, Song, Y.%
\BCBL {}\ \BBA {} Yang, S.%
\end{APACrefauthors}%
\unskip\
\newblock
\APACrefYearMonthDay{2020}{{\APACmonth{07}}}{}.
\newblock
{\BBOQ}\APACrefatitle {Gender disparities in the field of economics} {Gender
  disparities in the field of economics}.{\BBCQ}
\newblock
\APACjournalVolNumPages{Scientometrics}{125}{2}{1477--1498}.
\newblock
\begin{APACrefDOI} \doi{10.1007/s11192-020-03627-x} \end{APACrefDOI}
\PrintBackRefs{\CurrentBib}

\bibitem [\protect \citeauthoryear {%
Maddi%
\ \BBA {} Gingras%
}{%
Maddi%
\ \BBA {} Gingras%
}{%
{\protect \APACyear {2021}}%
}]{%
Maddi2021}
\APACinsertmetastar {%
Maddi2021}%
\begin{APACrefauthors}%
Maddi, A.%
\BCBT {}\ \BBA {} Gingras, Y.%
\end{APACrefauthors}%
\unskip\
\newblock
\APACrefYearMonthDay{2021}{{\APACmonth{02}}}{}.
\newblock
{\BBOQ}\APACrefatitle {GENDER DIVERSITY IN RESEARCH TEAMS AND CITATION IMPACT
  IN ECONOMICS AND MANAGEMENT} {Gender diversity in research teams and citation
  impact in economics and management}.{\BBCQ}
\newblock
\APACjournalVolNumPages{Journal of Economic Surveys}{35}{5}{1381--1404}.
\newblock
\begin{APACrefDOI} \doi{10.1111/joes.12420} \end{APACrefDOI}
\PrintBackRefs{\CurrentBib}

\bibitem [\protect \citeauthoryear {%
Mryglod%
}{%
Mryglod%
}{%
{\protect \APACyear {2012}}%
}]{%
Mryglod2012}
\APACinsertmetastar {%
Mryglod2012}%
\begin{APACrefauthors}%
Mryglod, O.%
\end{APACrefauthors}%
\unskip\
\newblock
\APACrefYearMonthDay{2012}{}{}.
\newblock
{\BBOQ}\APACrefatitle {Ukrainian scientific academic periodicals: The level of
  visibility.} {Ukrainian scientific academic periodicals: The level of
  visibility.}{\BBCQ}
\newblock
\APACjournalVolNumPages{Science of Ukraine in the Global Information
  Space}{}{6}{36--44}.
\PrintBackRefs{\CurrentBib}

\bibitem [\protect \citeauthoryear {%
Mryglod%
, Nazarovets%
\BCBL {}\ \BBA {} Kozmenko%
}{%
Mryglod%
\ \protect \BOthers {.}}{%
{\protect \APACyear {2021}}%
}]{%
Mryglod2021}
\APACinsertmetastar {%
Mryglod2021}%
\begin{APACrefauthors}%
Mryglod, O.%
, Nazarovets, S.%
\BCBL {}\ \BBA {} Kozmenko, S.%
\end{APACrefauthors}%
\unskip\
\newblock
\APACrefYearMonthDay{2021}{{\APACmonth{06}}}{}.
\newblock
{\BBOQ}\APACrefatitle {Universal and specific features of Ukrainian economic
  research: publication analysis based on Crossref data} {Universal and
  specific features of ukrainian economic research: publication analysis based
  on crossref data}.{\BBCQ}
\newblock
\APACjournalVolNumPages{Scientometrics}{126}{9}{8187--8203}.
\newblock
\begin{APACrefDOI} \doi{10.1007/s11192-021-04079-7} \end{APACrefDOI}
\PrintBackRefs{\CurrentBib}

\bibitem [\protect \citeauthoryear {%
M{\"u}ller%
, Reitz%
\BCBL {}\ \BBA {} Roy%
}{%
M{\"u}ller%
\ \protect \BOthers {.}}{%
{\protect \APACyear {2017}}%
}]{%
Muller2017}
\APACinsertmetastar {%
Muller2017}%
\begin{APACrefauthors}%
M{\"u}ller, M\BHBI C.%
, Reitz, F.%
\BCBL {}\ \BBA {} Roy, N.%
\end{APACrefauthors}%
\unskip\
\newblock
\APACrefYearMonthDay{2017}{}{}.
\newblock
{\BBOQ}\APACrefatitle {Data sets for author name disambiguation: an empirical
  analysis and a new resource} {Data sets for author name disambiguation: an
  empirical analysis and a new resource}.{\BBCQ}
\newblock
\APACjournalVolNumPages{Scientometrics}{111}{3}{1467--1500}.
\PrintBackRefs{\CurrentBib}

\bibitem [\protect \citeauthoryear {%
Nicola%
\ \BBA {} D'Agostino%
}{%
Nicola%
\ \BBA {} D'Agostino%
}{%
{\protect \APACyear {2021}}%
}]{%
DeNicola2021}
\APACinsertmetastar {%
DeNicola2021}%
\begin{APACrefauthors}%
Nicola, A\BPBI D.%
\BCBT {}\ \BBA {} D'Agostino, G.%
\end{APACrefauthors}%
\unskip\
\newblock
\APACrefYearMonthDay{2021}{{\APACmonth{03}}}{}.
\newblock
{\BBOQ}\APACrefatitle {Assessment of gender divide in scientific communities}
  {Assessment of gender divide in scientific communities}.{\BBCQ}
\newblock
\APACjournalVolNumPages{Scientometrics}{126}{5}{3807--3840}.
\newblock
\begin{APACrefDOI} \doi{10.1007/s11192-021-03885-3} \end{APACrefDOI}
\PrintBackRefs{\CurrentBib}

\bibitem [\protect \citeauthoryear {%
Reynolds%
\ \BBA {} Lewis%
}{%
Reynolds%
\ \BBA {} Lewis%
}{%
{\protect \APACyear {2017}}%
}]{%
Reynolds2017}
\APACinsertmetastar {%
Reynolds2017}%
\begin{APACrefauthors}%
Reynolds, A.%
\BCBT {}\ \BBA {} Lewis, D.%
\end{APACrefauthors}%
\unskip\
\newblock
\APACrefYearMonthDay{2017}{}{}.
\newblock
\APACrefbtitle {Teams Solve Problems Faster When They’re More Cognitively
  Diverse.} {Teams solve problems faster when they’re more cognitively
  diverse.}
\newblock
\APAChowpublished
  {\url{https://hbr.org/2017/03/teams-solve-problems-faster-when-theyre-more-cognitively-diverse}}.
\newblock
\APACrefnote{[Online; accessed 05-July-2022]}
\PrintBackRefs{\CurrentBib}

\bibitem [\protect \citeauthoryear {%
Sasv{\'{a}}ri%
, Nemeslaki%
\BCBL {}\ \BBA {} Duma%
}{%
Sasv{\'{a}}ri%
\ \protect \BOthers {.}}{%
{\protect \APACyear {2019}}%
}]{%
Sasvri2019}
\APACinsertmetastar {%
Sasvri2019}%
\begin{APACrefauthors}%
Sasv{\'{a}}ri, P.%
, Nemeslaki, A.%
\BCBL {}\ \BBA {} Duma, L.%
\end{APACrefauthors}%
\unskip\
\newblock
\APACrefYearMonthDay{2019}{{\APACmonth{03}}}{}.
\newblock
{\BBOQ}\APACrefatitle {Exploring the influence of scientific journal ranking on
  publication performance in the Hungarian social sciences: the case of law and
  economics} {Exploring the influence of scientific journal ranking on
  publication performance in the hungarian social sciences: the case of law and
  economics}.{\BBCQ}
\newblock
\APACjournalVolNumPages{Scientometrics}{119}{2}{595--616}.
\newblock
\begin{APACrefDOI} \doi{10.1007/s11192-019-03081-4} \end{APACrefDOI}
\PrintBackRefs{\CurrentBib}

\bibitem [\protect \citeauthoryear {%
Schl\"{a}pfer%
}{%
Schl\"{a}pfer%
}{%
{\protect \APACyear {2010}}%
}]{%
Schlapfer2010}
\APACinsertmetastar {%
Schlapfer2010}%
\begin{APACrefauthors}%
Schl\"{a}pfer, F.%
\end{APACrefauthors}%
\unskip\
\newblock
\APACrefYearMonthDay{2010}{{\APACmonth{06}}}{}.
\newblock
{\BBOQ}\APACrefatitle {How Much Does Journal Reputation Tell Us About the
  Academic Interest and Relevance of Economic Research? Empirical Analysis and
  Implications for Environmental Economic Research} {How much does journal
  reputation tell us about the academic interest and relevance of economic
  research? empirical analysis and implications for environmental economic
  research}.{\BBCQ}
\newblock
\APACjournalVolNumPages{{GAIA} - Ecological Perspectives for Science and
  Society}{19}{2}{140--145}.
\newblock
\begin{APACrefDOI} \doi{10.14512/gaia.19.2.13} \end{APACrefDOI}
\PrintBackRefs{\CurrentBib}

\bibitem [\protect \citeauthoryear {%
Thelwall%
, Bailey%
, Tobin%
\BCBL {}\ \BBA {} Bradshaw%
}{%
Thelwall%
\ \protect \BOthers {.}}{%
{\protect \APACyear {2019}}%
}]{%
Thelwall2019}
\APACinsertmetastar {%
Thelwall2019}%
\begin{APACrefauthors}%
Thelwall, M.%
, Bailey, C.%
, Tobin, C.%
\BCBL {}\ \BBA {} Bradshaw, N\BHBI A.%
\end{APACrefauthors}%
\unskip\
\newblock
\APACrefYearMonthDay{2019}{{\APACmonth{02}}}{}.
\newblock
{\BBOQ}\APACrefatitle {Gender differences in research areas, methods and
  topics: Can people and thing orientations explain the results?} {Gender
  differences in research areas, methods and topics: Can people and thing
  orientations explain the results?}{\BBCQ}
\newblock
\APACjournalVolNumPages{Journal of Informetrics}{13}{1}{149--169}.
\newblock
\begin{APACrefDOI} \doi{10.1016/j.joi.2018.12.002} \end{APACrefDOI}
\PrintBackRefs{\CurrentBib}

\bibitem [\protect \citeauthoryear {%
Thelwall%
\ \BBA {} Mas-Bleda%
}{%
Thelwall%
\ \BBA {} Mas-Bleda%
}{%
{\protect \APACyear {2020}}%
}]{%
Thelwall2020}
\APACinsertmetastar {%
Thelwall2020}%
\begin{APACrefauthors}%
Thelwall, M.%
\BCBT {}\ \BBA {} Mas-Bleda, A.%
\end{APACrefauthors}%
\unskip\
\newblock
\APACrefYearMonthDay{2020}{{\APACmonth{08}}}{}.
\newblock
{\BBOQ}\APACrefatitle {A gender equality paradox in academic publishing:
  Countries with a higher proportion of female first-authored journal articles
  have larger first-author gender disparities between fields} {A gender
  equality paradox in academic publishing: Countries with a higher proportion
  of female first-authored journal articles have larger first-author gender
  disparities between fields}.{\BBCQ}
\newblock
\APACjournalVolNumPages{Quantitative Science Studies}{1}{3}{1260--1282}.
\newblock
\begin{APACrefDOI} \doi{10.1162/qss_a_00050} \end{APACrefDOI}
\PrintBackRefs{\CurrentBib}

\bibitem [\protect \citeauthoryear {%
Treeratpituk%
\ \BBA {} Giles%
}{%
Treeratpituk%
\ \BBA {} Giles%
}{%
{\protect \APACyear {2012}}%
}]{%
Treeratpituk2012}
\APACinsertmetastar {%
Treeratpituk2012}%
\begin{APACrefauthors}%
Treeratpituk, P.%
\BCBT {}\ \BBA {} Giles, C\BPBI L.%
\end{APACrefauthors}%
\unskip\
\newblock
\APACrefYearMonthDay{2012}{}{}.
\newblock
{\BBOQ}\APACrefatitle {Name-ethnicity classification and ethnicity-sensitive
  name matching} {Name-ethnicity classification and ethnicity-sensitive name
  matching}.{\BBCQ}
\newblock
\BIn{} \APACrefbtitle {Proceedings of the AAAI Conference on Artificial
  Intelligence} {Proceedings of the aaai conference on artificial
  intelligence}\ (\BVOL~26, \BPGS\ 1141--1147).
\PrintBackRefs{\CurrentBib}

\bibitem [\protect \citeauthoryear {%
Truc%
, Claveau%
\BCBL {}\ \BBA {} Santerre%
}{%
Truc%
\ \protect \BOthers {.}}{%
{\protect \APACyear {2021}}%
}]{%
Truc2021}
\APACinsertmetastar {%
Truc2021}%
\begin{APACrefauthors}%
Truc, A.%
, Claveau, F.%
\BCBL {}\ \BBA {} Santerre, O.%
\end{APACrefauthors}%
\unskip\
\newblock
\APACrefYearMonthDay{2021}{{\APACmonth{01}}}{}.
\newblock
{\BBOQ}\APACrefatitle {Economic methodology: a bibliometric perspective}
  {Economic methodology: a bibliometric perspective}.{\BBCQ}
\newblock
\APACjournalVolNumPages{Journal of Economic Methodology}{28}{1}{67--78}.
\newblock
\begin{APACrefDOI} \doi{10.1080/1350178x.2020.1868774} \end{APACrefDOI}
\PrintBackRefs{\CurrentBib}

\bibitem [\protect \citeauthoryear {%
Vaio%
\ \BBA {} Weisdorf%
}{%
Vaio%
\ \BBA {} Weisdorf%
}{%
{\protect \APACyear {2009}}%
}]{%
DiVaio2009}
\APACinsertmetastar {%
DiVaio2009}%
\begin{APACrefauthors}%
Vaio, G\BPBI D.%
\BCBT {}\ \BBA {} Weisdorf, J\BPBI L.%
\end{APACrefauthors}%
\unskip\
\newblock
\APACrefYearMonthDay{2009}{{\APACmonth{03}}}{}.
\newblock
{\BBOQ}\APACrefatitle {Ranking economic history journals: a citation-based
  impact-adjusted analysis} {Ranking economic history journals: a
  citation-based impact-adjusted analysis}.{\BBCQ}
\newblock
\APACjournalVolNumPages{Cliometrica}{4}{1}{1--17}.
\newblock
\begin{APACrefDOI} \doi{10.1007/s11698-009-0039-y} \end{APACrefDOI}
\PrintBackRefs{\CurrentBib}

\bibitem [\protect \citeauthoryear {%
Waltman%
}{%
Waltman%
}{%
{\protect \APACyear {2012}}%
}]{%
Waltman2012}
\APACinsertmetastar {%
Waltman2012}%
\begin{APACrefauthors}%
Waltman, L.%
\end{APACrefauthors}%
\unskip\
\newblock
\APACrefYearMonthDay{2012}{{\APACmonth{10}}}{}.
\newblock
{\BBOQ}\APACrefatitle {An empirical analysis of the use of alphabetical
  authorship in scientific publishing} {An empirical analysis of the use of
  alphabetical authorship in scientific publishing}.{\BBCQ}
\newblock
\APACjournalVolNumPages{Journal of Informetrics}{6}{4}{700--711}.
\newblock
\begin{APACrefDOI} \doi{10.1016/j.joi.2012.07.008} \end{APACrefDOI}
\PrintBackRefs{\CurrentBib}

\bibitem [\protect \citeauthoryear {%
Wei%
}{%
Wei%
}{%
{\protect \APACyear {2018}}%
}]{%
Wei2018}
\APACinsertmetastar {%
Wei2018}%
\begin{APACrefauthors}%
Wei, G.%
\end{APACrefauthors}%
\unskip\
\newblock
\APACrefYearMonthDay{2018}{{\APACmonth{03}}}{}.
\newblock
{\BBOQ}\APACrefatitle {A BIBLIOMETRIC ANALYSIS OF THE TOP FIVE ECONOMICS
  JOURNALS DURING 2012-2016} {A bibliometric analysis of the top five economics
  journals during 2012-2016}.{\BBCQ}
\newblock
\APACjournalVolNumPages{Journal of Economic Surveys}{33}{1}{25--59}.
\newblock
\begin{APACrefDOI} \doi{10.1111/joes.12260} \end{APACrefDOI}
\PrintBackRefs{\CurrentBib}

\bibitem [\protect \citeauthoryear {%
West%
, Jacquet%
, King%
, Correll%
\BCBL {}\ \BBA {} Bergstrom%
}{%
West%
\ \protect \BOthers {.}}{%
{\protect \APACyear {2013}}%
}]{%
West2013}
\APACinsertmetastar {%
West2013}%
\begin{APACrefauthors}%
West, J\BPBI D.%
, Jacquet, J.%
, King, M\BPBI M.%
, Correll, S\BPBI J.%
\BCBL {}\ \BBA {} Bergstrom, C\BPBI T.%
\end{APACrefauthors}%
\unskip\
\newblock
\APACrefYearMonthDay{2013}{{\APACmonth{07}}}{}.
\newblock
{\BBOQ}\APACrefatitle {The Role of Gender in Scholarly Authorship} {The role of
  gender in scholarly authorship}.{\BBCQ}
\newblock
\APACjournalVolNumPages{{PLoS} {ONE}}{8}{7}{e66212}.
\newblock
\begin{APACrefDOI} \doi{10.1371/journal.pone.0066212} \end{APACrefDOI}
\PrintBackRefs{\CurrentBib}

\bibitem [\protect \citeauthoryear {%
Wohlrabe%
\ \BBA {} Bornmann%
}{%
Wohlrabe%
\ \BBA {} Bornmann%
}{%
{\protect \APACyear {2022}}%
}]{%
Wohlrabe2022}
\APACinsertmetastar {%
Wohlrabe2022}%
\begin{APACrefauthors}%
Wohlrabe, K.%
\BCBT {}\ \BBA {} Bornmann, L.%
\end{APACrefauthors}%
\unskip\
\newblock
\APACrefYearMonthDay{2022}{{\APACmonth{03}}}{}.
\newblock
{\BBOQ}\APACrefatitle {Alphabetized co-authorship in economics reconsidered}
  {Alphabetized co-authorship in economics reconsidered}.{\BBCQ}
\newblock
\APACjournalVolNumPages{Scientometrics}{127}{5}{2173--2193}.
\newblock
\begin{APACrefDOI} \doi{10.1007/s11192-022-04322-9} \end{APACrefDOI}
\PrintBackRefs{\CurrentBib}

\bibitem [\protect \citeauthoryear {%
Zhao%
, Yu%
, Tan%
, Xu%
\BCBL {}\ \BBA {} Yu%
}{%
Zhao%
\ \protect \BOthers {.}}{%
{\protect \APACyear {2016}}%
}]{%
Zhao2016}
\APACinsertmetastar {%
Zhao2016}%
\begin{APACrefauthors}%
Zhao, S\BPBI X.%
, Yu, S.%
, Tan, A\BPBI M.%
, Xu, X.%
\BCBL {}\ \BBA {} Yu, H.%
\end{APACrefauthors}%
\unskip\
\newblock
\APACrefYearMonthDay{2016}{{\APACmonth{05}}}{}.
\newblock
{\BBOQ}\APACrefatitle {Global pattern of science funding in economics} {Global
  pattern of science funding in economics}.{\BBCQ}
\newblock
\APACjournalVolNumPages{Scientometrics}{109}{1}{463--479}.
\newblock
\begin{APACrefDOI} \doi{10.1007/s11192-016-1961-y} \end{APACrefDOI}
\PrintBackRefs{\CurrentBib}

\end{thebibliography}
\end{document}